\documentclass[a4paper,fleqn]{cas-sc}

\usepackage{multirow}

\def\tsc#1{\csdef{#1}{\textsc{\lowercase{#1}}\xspace}}
\tsc{WGM}
\tsc{QE}
\tsc{EP}
\tsc{PMS}
\tsc{BEC}
\tsc{DE}

\usepackage{soul}

\usepackage{pmboxdraw}
\usepackage{lscape}
\usepackage[dvipsnames]{xcolor}
\usepackage{algorithm}
\usepackage{algorithmic}
\usepackage{caption}
\usepackage{changepage}
\usepackage{tabularx}
\usepackage{array}
\usepackage[square,sort,comma,numbers]{natbib}
\usepackage{mathtools}

\usepackage{lipsum}
\makeatletter
\def\ps@pprintTitle{%
 \let\@oddhead\@empty
 \let\@evenhead\@empty
 \def\@oddfoot{}%
 \let\@evenfoot\@oddfoot}
\makeatother

\DeclarePairedDelimiter\ket{\lvert}{\rangle}

\setcitestyle{square}
\begin{document}
\let\WriteBookmarks\relax
\def\floatpagepagefraction{1}
\def\textpagefraction{.001}

\shorttitle{Quantum Computing: Vision and Challenges}
\shortauthors{Cite}

\font\myfont=cmr12 at 23pt \title{{\myfont Quantum Computing: Vision and Challenges}}

\author[1]{Sukhpal Singh Gill} [orcid=0000-0002-3913-0369]
\ead{s.s.gill@qmul.ac.uk}
\affiliation[1]{organization={School of Electronic Engineering and Computer Science, Queen Mary University of London, London, UK}}

\author[2]{Oktay Cetinkaya}
\affiliation[2]{organization={Oxford e-Research Centre (OeRC), Department of Engineering Science, University of Oxford, Oxford, UK}}
\ead{oktay.cetinkaya@eng.ox.ac.uk}

\author[3]{Stefano Marrone}
\affiliation[3]{organization={Dipartimento di Matematica e Fisica, Università della Campania “Luigi Vanvitelli”, Italy}}
\ead{stefano.marrone@unicampania.it}

\author[4]{Daniel Claudino}
\affiliation[4]{organization={Quantum Information Science Section, Oak Ridge National Laboratory, Oak Ridge, TN, USA}}
\ead{claudinodc@ornl.gov}

\author[5]{David Haunschild}
\affiliation[5]{organization={Detecon International GmbH, Munich, Germany}}
\ead{david.haunschild@detecon.com}

\author[6]{Leon Schlote}
\affiliation[6]{organization={DB Cargo, Berlin, Germany}}
\ead{leon.schlote@deutschebahn.com}

\author[7]{Huaming Wu}
\affiliation[7]{organization={Center for Applied Mathematics, Tianjin University, Tianjin, China}}
\ead{whming@tju.edu.cn}

\author[8]{Carlo Ottaviani}
\affiliation[8]{organization={Department of Computer Science and York Centre for Quantum Technologies, University of York, York, UK}}
\ead{carlo.ottaviani@york.ac.uk}

\author[9]{Xiaoyuan Liu}
\affiliation[9]{organization={Quantum Laboratory, Fujitsu Research of America, Inc., Santa Clara, CA, USA}}
\ead{xliu@fujitsu.com}

\author[10]{Sree Pragna Machupalli}
\affiliation[10]{organization={Information Networking Institute, Carnegie Mellon University, Pittsburgh, Pennsylvania, USA}}
\ead{smachupa@andrew.cmu.edu}

\author[11]{Kamalpreet Kaur}
\affiliation[11]{organization={Cymax Group Technologies, Burnaby, British Columbia, Canada}}
\ead{kamalpreet.k@cymax.com}

\author[12]{Priyansh Arora}
\affiliation[12]{organization={Microsoft, Schiphol, Netherlands}}
\ead{priyansh.arora@microsoft.com}

\author[13]{Ji Liu}
\affiliation[13]{organization={Mathematics and Computational Research Division, Argonne National Laboratory, IL, USA}}
\ead{ji.liu@anl.gov}

\author[14]{Ahmed Farouk}
\affiliation[14]{organization={Department of Computer Science, Faculty of Computers and Artificial Intelligence, South Valley University, Hurghada, Egypt}}
\ead{ahmed.farouk@sci.svu.edu.eg}

\author[15]{Houbing Herbert Song}
\affiliation[15]{organization={Department of Information Systems University of Maryland, Baltimore County (UMBC), Baltimore, USA}}
\ead{songh@umbc.edu}

\author[1]{Steve Uhlig}
\ead{steve.uhlig@qmul.ac.uk}

\author[16]{Kotagiri Ramamohanarao}
\affiliation[16]{organization={Retired Professor, The University of Melbourne, Melbourne, Victoria, Australia}}
\ead{rkotagiri@gmail.com}

\cortext[cor2]{Correspondence to: School of Electronic Engineering and Computer Science, Queen Mary University of London, London, E1 4NS, UK.}

\fntext[fn1]{This is the DOI of a published book chapter on Elsevier: \url{https://doi.org/10.1016/B978-0-443-29096-1.00008-8}}

\begin{abstract} 
The recent development of quantum computing, which uses entanglement, superposition, and other quantum fundamental concepts, can provide substantial processing advantages over traditional computing. These quantum features help solve many complex problems that cannot be solved otherwise with conventional computing methods. These problems include modeling quantum mechanics, logistics, chemical-based advances, drug design, statistical science, sustainable energy, banking, reliable communication, and quantum chemical engineering. The last few years have witnessed remarkable progress in quantum software and algorithm creation and quantum hardware research, which has significantly advanced the prospect of realizing quantum computers. It would be helpful to have comprehensive literature research on this area to grasp the current status and find outstanding problems that require considerable attention from the research community working in the quantum computing industry. To better understand quantum computing, this paper examines the foundations and vision based on current research in this area. We discuss cutting-edge developments in quantum computer hardware advancement and subsequent advances in quantum cryptography, quantum software, and high-scalability quantum computers. Many potential challenges and exciting new trends for quantum technology research and development are highlighted in this paper for a broader debate.
\end{abstract}

\maketitle

\section{Promising Age of Quantum Computing}

Many experts regard Richard Feynman's 1982 talk as among the first ideas for quantum computing \cite{hey1999richard,preskill2023quantum}. Feynman imagined a quantum machine that could imitate quantum physics by using the principles of quantum mechanics. According to Feynman's view, a computer based on quantum mechanical fundamentals might be necessary to mimic natural occurrences, as Nature is fundamentally quantum mechanical \cite{silva2023richard}. The advent of quantum computers has opened up new avenues for this kind of thinking, since they can harness the incredible processing power required to model intricate quantum systems by making use of quantum mechanical features such as superposition, interference, and entanglement \cite{yang2023survey}. Early efforts to build hardware for quantum computers moved at a \emph{snail's pace} due to challenging technical problems, making it difficult to shield and coherently control the dynamics of quantum mechanical properties present at the most essential scales of nature (e.g., electron spin or photon polarization) \cite{mikkelsen2007optically}. 

However, quantum computing is one of the most talked-about fields  (as of 2024), and its progress has been growing at a tremendous pace in recent years \cite{gill2024}. There is a great deal of enthusiasm among academics and businesses alike to construct initial quantum computers due to their promise of providing, for certain tasks, processing powers beyond those of current most powerful supercomputers. Strong efforts to build large-scale quantum computers are now underway with several established corporations (Chinese companies like ZTE, QUDOOR and USA based companies such as Honeywell, Intel, Google, Microsoft, and IBM), growing small and medium-sized enterprises (e.g., D-Wave), and aspiring startups (e.g., Rigetti, Xanadu, Infleqtion, Origin Quantum, and IonQ). There has been enormous advancement in quantum algorithms and quantum software in recent years, which has occurred in tandem with the development of quantum hardware. 

It is well-known that traditional digital computing relies on bits that are limited to two possible values—`0' or `1'—to store and process data. In quantum computing, the corresponding unit is the quantum bit (qubit) that, according to quantum physics, may have either a value of `0' or `1' or exist on a superposition of the two (functionally being in both states simultaneously!) \cite{nielsen2010quantum,nadj2010spin,hendrickx2020single}. Because of this, quantum computers have access to a computational field (known as Hilbert space \cite{vourdas2004quantum}) of huge dimension, where $n$ qubits might be in a superposition state with $2^{n}$ potential values at any one moment. Due to the exponential growth of the parameter space, problems on a large scale are expected to be easier to solve with the advent of quantum computers. Nevertheless, developing a large-scale quantum computer has its own set of challenges. The most demanding to mitigate is the decoherence of the quantum states on which qubits are encoded. Decoherence happens when qubits interact with their surrounding environment and lose their coherent features. For that it represents one of the biggest obstacles to developing large-scale quantum devices \cite{kumar2022securing}. Assuming the unavoidable presence of environmental noise, ``Noisy Intermediate Scale Quantum (NISQ)'' devices, try to deal with imperfections and losses driven by decoherence. Reducing the probability of decoherence and creating effective error correction procedures to overcome defects in NISQ devices are important goals of current studies in quantum computing \cite{preskill2018quantum}. The second big problem with modern quantum devices is to identify approaches to effectively engineer and interconnect (entangle) qubits \cite{howard2023implementing}. At the moment of writing current quantum devices are able to deal with relatively sparsely connected qubits,  making it difficult to map deep quantum circuits with multiple two-qubit gates that necessitate strong couplings between qubits \cite{abughanem2024two}.

\textbf{1.1 Quantum Supremacy:} Regardless of technological hurdles, NISQ quantum computers have shown promising computing capability in their early stages. Google's recent proof of quantum supremacy is a major step forward for quantum computing \cite{arute2019quantum}. There is currently a worldwide race to be the first to implement quantum computing in order to tackle a practical problem that a conventional computer cannot solve in a reasonable time --- also known as \textit{``quantum advantage''}. To reach this desired level of quantum computing, it is necessary to reduce the probability of the decoherence of qubits drastically through improvements in quantum hardware, quantum algorithms, and error correction during the upcoming years. A lot of work is being put into developing and benchmarking quantum algorithms using NISQ devices. While  Shor's and Grover's quantum algorithms were among the first that stood out in the early 1990s, hundreds of other algorithms have been invented since then. Variational Quantum Eigensolver (VQE) \cite{peruzzo2014variational,kandala2017hardware} and other variational quantum algorithms \cite{cerezo2021variational} are a popular kind of hybrid quantum-classical algorithm that combines the advantages of the two technologies. On NISQ devices, VQE algorithms have performed exceptionally well in solving quantum mechanical problems and Quantum Artificial Intelligence (QAI) tasks \cite{singh2022quantum}. While a large and resilient quantum computer is not available yet and will still require significant advancements before its full promise for practical applications can be realised, quantum computing is already available for research and prototyping scenarios with encouraging results on current NISQ-era equipment \cite{corcoles2019challenges}.

\begin{figure*}
	\centering
	\includegraphics[width=1.05\textwidth]{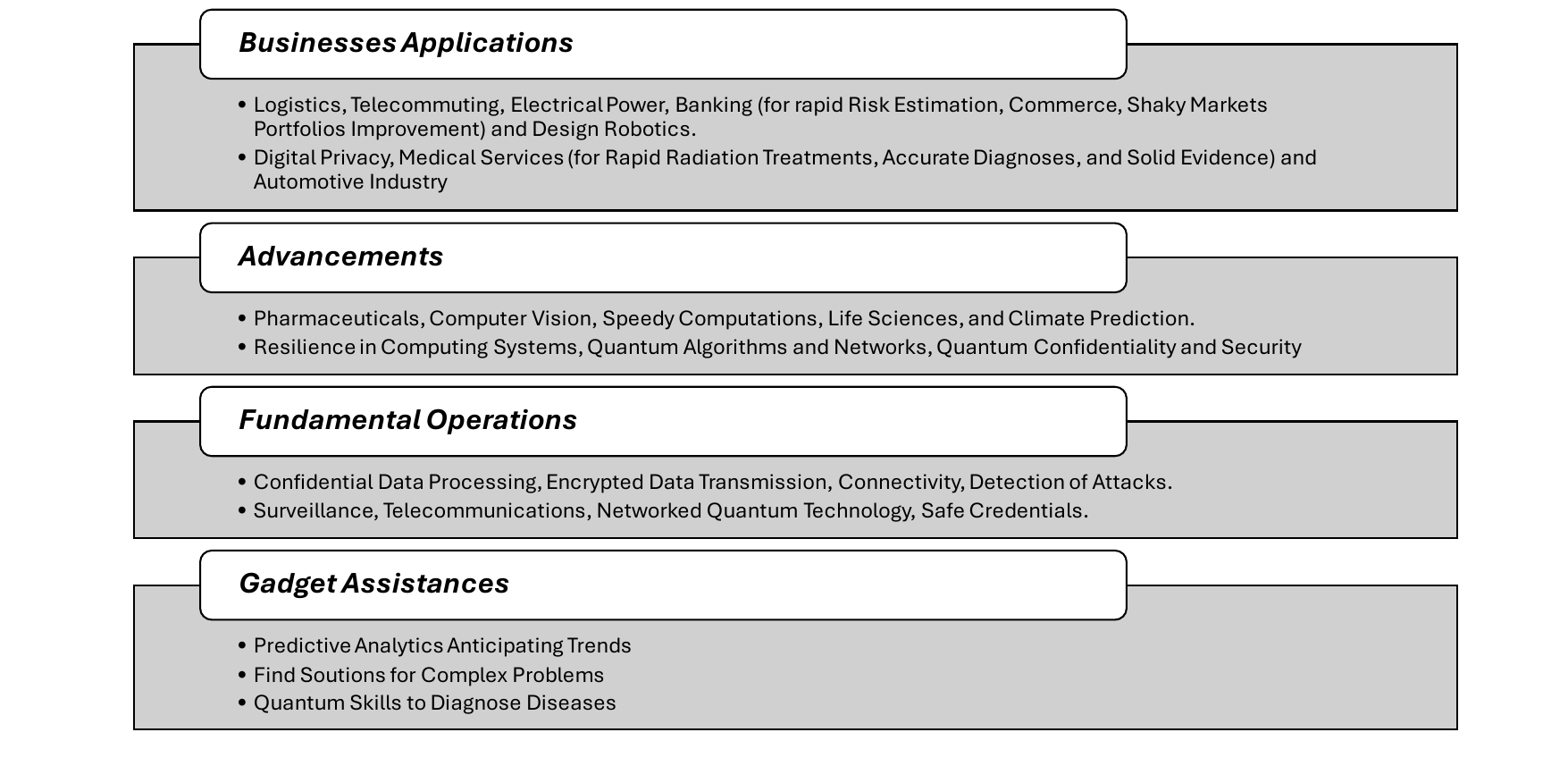}
	\caption [Caption for LOF] {Applications and Benefits of Quantum Computing.}
	\label{Applications}
\end{figure*}

When applied to classical data, QAI has the potential to greatly accelerate machine intelligence techniques \cite{krenn2023artificial, biamonte2017quantum}. Quantum neural networks, quantum support vector machines, and quantum principle component evaluation have been studied \cite{mafu2021design, rebentrost2014quantum}, and some recent research returned encouraging findings \cite{ding2021quantum}, although it is still not completely known if quantum neural networks will provide better computing efficiency than traditional machine learning techniques.

There exist several different quantum computing paradigms. The most popular ones are measurement-based or one-way quantum computing \cite{browne2016one}, adiabatic quantum computing (usually implemented in practice as quantum annealing) \cite{albash2018adiabatic}, and the quantum circuit framework for gate-based general quantum computing \cite{nielsen2010quantum}. Since it is possible to re-program quantum computers according to particular issues, the quantum circuit model stands out as an especially feasible option. Currently, some high-level programming languages specific to quantum computing, such as Qiskit \cite{cross2018ibm}, Cirq \cite{heim2020quantum}, PennyLane \cite{bergholm2018pennylane}, and other libraries and packages, are available to program quantum computers; however, circuits specified with these languages need to be ``translated'' to fit the actual quantum topology, building the quantum circuits by organising the necessary quantum gates (these are just ``instructions'' that are executed in sequence) and operations according to a pre-designed architecture.

\textbf{1.2 Applications and Benefits:} Research on quantum computing is blossoming, with regular exciting new advances in several areas of application and quantum engineering such as hardware, software, algorithms, error correction on NISQ devices. Academic scientists first, but now also industry experts are investigating on problems that may find applications to solve practical problems. In Fig.~\ref{Applications}, we summarize some benefits that quantum computing may have for common users, programmers, and various business sectors by delegating key tasks.

\textbf{1.3 Quantum Computing in a nutshell:} A binary bit that may take on values `0' or `1' is the basic unit of information of conventional computing. Quantum Computation and Information uses qubits as fundamental unit of information and, differently from classical bits, they can not only acquire either value `0' or `1', but even `0' and `1' at the same time. A simple mathematical representation of a qubit, in the computational basis $\{|0\rangle, |1\rangle\}$, is conventionally given as: 
\begin{equation}
    a\ket{0} + b\ket{1},
\end{equation} 
where $a$ and $b$ are complex amplitudes ($a, b \in \mathcal{C}$) superimposing the states `0' and `1' \cite{preskill2023quantum}, and preserving probability interpretation of quantum state, i.e., they need to verify the condition $|a|^2+|b|^2=1$. The symbol $|\rangle$ (ket) indicates that the bit of information is encoded in a quantum state, exploiting one of its physical degree of freedom. Using quantum superposition, a vast computational space becomes available allowing to solve problems of extreme complexity \cite{nielsen2010quantum}. Even a very limited number of qubits, $N$ can be used to solve problems that are intractable with classical computers, thanks to the rapidly expanding computational domain as an exponential function ($2^N$) of the total number of qubits. 

Another fundamental quantum property exploited in quantum algorithms is \textit{Entanglement}  \cite{nielsen2010quantum}. While classical bits are independent of each other when setting bit values, qubits allow for the placement of bits in an entangled state. When entangled qubits can persist in a correlated global state, even if physically apart. As a result, all qubits in an entangled state can have their characteristics changed even if only one of them is probed. When used for dense coding or quantum simulation of linked networks, entanglement becomes a valuable asset \cite{gill2021quantum}. 

Measurement is the last stage of a quantum computation; it collapses the stochastic quantum state into a deterministic state. Although quantum algorithms typically guarantee that the correct outcome has the highest likelihood, the stochastic nature of the process cannot guarantee that the correct outcome is actually sampled. Therefore, some classical post-processing (such as majority voting or statistical estimation) or repeating the computation several times is usually needed to produce a final output from the raw results obtained with the quantum computer.

\section{Quantum Algorithms}
A quantum computer is based on the principles of quantum mechanics and uses these principles to its advantage. From their origins in quantum physics models to many modern computer science uses, quantum algorithms have come a long way \cite{preskill2023quantum}. A highly coveted step towards attaining the processing capacity of its type, an industrial-scale quantum computer would certainly have ramifications in several domains, including cybersecurity and others. Daniel Simon presented the first quantum method to beat classical algorithms in terms of performance \cite{simon1997power}. Deutsch-Jozsa Algorithm, Bernstein-Vazirani Algorithm, Simon’s Algorithm, and Shor’s Algorithm were introduced to focus on problems that require exponential queries  (i.e. cutting down on the amount of computing power required to examine algorithms and assess their balance or robustness with certainty), efficient solutions of black-box problems, faster computation, speedup, and integer factorisation, and discrete logarithm problems, respectively \cite{yang2023survey}. These algorithms were based on the quantum Fourier transform. Furthermore, Grover's algorithm and quantum counting were developed to concentrate on searching unstructured databases for marked entries and generalised searches, respectively. Both of these algorithms were created based on amplitude amplification, which is a robust strategy to make quantum computers capable of solving challenges quickly and effectively that might be impossible to solve with traditional approaches. Numerous quantum algorithms rely on this, such as those for quantum machine learning, quantum simulation, and quantum search. Finally, a quantum approximate optimisation approach centered on the solution of graph theory issues has been recently proposed \cite{farhi2014quantum}. This approach is built on a hybrid quantum/classical scheme. 
From a foundational point of view, all software-related aspects are based on two different computational models, which determine some differences in the programming paradigms as well as in applications and technical aspects: the quantum gate \cite{Williams2011} and quantum annealing models \cite{Du20081501}. The gate model uses quantum gates to perform operations on qubits. These gates manipulate qubits in a manner similar to classical logic gates, with the ability to exploit quantum-related features such as entanglement and superposition. This is a universal computational model in which the above-mentioned Shor's and Grover's algorithms can be implemented; hence, the applications based on this model have the widest range. From a technical point of view, decoherence is the main problem, and error correction is the most required practice. On the other hand, quantum annealing is an approximate implementation of adiabatic quantum computing, which is itself equivalent to the digital model, which seems to be a promising alternative to the gate model for solving large optimisation problems. This paradigm is based on the natural tendency of quantum systems to find low-energy states. It relies on the natural quantum mechanical process of tunneling and requires maintaining a coherent quantum state over the annealing process. It is somewhat less sensitive to errors compared to the gate model because it exploits the quantum system's natural tendency to find a low-energy state, making it robust against certain types of computational errors.

\section{Technological Advances and Software Tools}
The invention of quantum software is an emerging yet relatively less developed field compared to quantum modelling and quantum technology \cite{de2022software}. Several quantum applications are already accessible from various platforms/sources, including Google, IBM, Microsoft, and D-Wave. Quantum programming tools have been produced at an increasing pace; however, there is a lack of excellent programming tools, similar to conventional programming languages like C++ and Java, and these applications are still rather low-level, like assembly-level languages. A number of important areas have been identified in recent research pertaining to software programs that use quantum computing, including coding languages, programmers, error-correction firmware, physical level schedulers and optimisers, logical level schedulers and optimisation techniques, and hardware control of software updates.  Most important topics to study in the field are \cite{serrano2022quantum, perez2021software, vietz2021decision}: (i) frameworks, semantics and compilation of programming language; (ii) workflows, controlled and adjoint operations \& clean and borrowed qubits and (iii) simulators. Effectively integrating quantum algorithms with defective equipment is the goal of powerful quantum error-correcting firmware \cite{serrano2022quantum}. Located at the very bottom of the quantum computing stack, error-correcting quantum firmware aids in lowering the error rate due to flawed hardware, as well as the intricacy and resource consumption of the system \cite{perez2021software}. It is envisaged that software managing quantum hardware would have outstanding performance, be able to use sophisticated quantum management techniques, have top-quality effects at the system level, be able to regulate for both global and local optimal outcomes through simulation, and have adequate physical schedules \cite{vietz2021decision}.
At this date, notwithstanding the absence of a single programming framework/model able to overcome the others, there are different platforms for quantum computer programming, often provided and ``tied'' to the provided hardware solutions. Among the most famous are: Qiskit (Quantum Information Science Kit) --- developed by IBM\footnote{https://www.ibm.com/quantum/qiskit}; Cirq --- developed by Google\footnote{https://quantumai.google/cirq}; and PyQuil --- developed by Rigetti Computing\footnote{https://github.com/rigetti/pyquil}. To push their solutions, quantum developers often release these frameworks with open-source licenses and with an Application Programming Interface (API) in Python, which is a language that is straightforward to learn. Quantum Annealing is following the same path, with a couple of “programming frameworks” --- i.e., D-Wave Ocean Software and Leap --- both provided by D-Wave. Recently, Fujitsu's Digital Annealer has been promising to bring quantum-inspired technology using traditional computing platforms \cite{Aramon2019}.

\section{Modern Cryptography: From Quantum to Post-Quantum}
The advent of quantum computers heralds a new ground-breaking era within the realm of data integrity and cybersecurity. With improving scalable computing power, quantum computers can effortlessly break the security of traditional cryptosystems, relying on factorization and discrete logarithms, both of which are considered hard problems for classical computers. By constrast, quantum computers have efficient processing capabilities to solve these hard problems within polynomial time~\cite{singh2021quantum}. For example, an adversary equipped with a quantum computer may break the RSA (Rivest-Shamir-Adleman) security in polynomial time by exploiting Shor’s algorithm for factoring large numbers. It is clear that such a possibility, despite not yet practical, poses potential threats to the integrity of communication networks ~\cite{shor1999polynomial} that need to be analyzed and mitigated. In fact the potential threat represented by the Shor's algorithm has led to new developments in classical cryptographic approaches, with the work on post-quantum cryptography and on a completely new paradigm to grant security named quantum cryptography ~\cite{pirandolaAQC}, or more precisely Quantum-Key Distribution (QKD). The novelty of QKD is that, instead of adding layers of security based on conventional (i.e. computationally hard to solve) algorithms, it uses fundamental properties of quantum particles to protect information from unauthorized parties.
QKD protocols, are themselves \emph{composite} algorithms where  
transmission of quantum signals, encryption/decryption, signatures, authentication, and hashing are all combined ~\cite{pirandola2016physics} to achieve (theoretically) unconditional security.

Let's revise in some more detail the basic principle of both quantum cryptography and post-quantum cryptography.

\textbf{4.1 Quantum Key Distribution:} Classical cryptography is endangered by the discovery of the Shor’s algorithm because it can efficiently solve computationally hard problems upon which classical key-exchange mechanism are based.
By contrast Quantum Key Distribution (QKD) does not make use of  computationally hard primitives, but relies on the fundamental laws of quantum physics to establish security. However, it is worth to notice that QKD protocols are always \emph{hybrid}, i.e., they rely on both quantum and classical communications to implement a virtually impenetrable crypto-system, promising to protect the privacy of communication even against attacks conducted by any quantum computers, independently from their computational power and evolution.

QKD can be implemented in two specific setups: continuous-variable (CV-QKD) and discrete-variable (DV-QKD) \cite{pirandolaAQC}. DV-QKD uses qubits to encode information and single photon detectors are employed by the receiver to monitor and quantify the presence of eavesdropper on the communication channel \cite{zhang2019integrated}. In such a way the parties can quantify the amount of information eavesdropped. By contrast, CV-QKD encodes classical information randomly modulating phase and amplitude of bright coherent states, and uses homodyne detection schemes at the receivers, in a similar setup used today by conventional optical communications  \cite{matsuura2021finite}.

An essential tenet of QKD (both for DV and CV) is rooted in quantum physics and takes the shape of the no-cloning theorem~\cite{park-nocloning,zurek-nocloning}, which asserts that a flawless replica of arbitrary (i.e. non-orthogonal) quantum states cannot be created, without corrupting the probed quantum states. That is exploited during the \emph{quantum communication phase}, when quantum particles are sent from the sender to the receiver. In fact, encoding  information on non-orthogonal quantum states, makes so that any effort to gain insights on the properties of such a stream quantum signals would results in the introduction of noise, readily identified by either the key distributor or the recipient (the parties, conventionally Alice and Bob). Such a mechanism allows the parties to quantify the amount of information potentially eavesdropped (Eve) during the quantum communication. That information is crucial, because they can use it to then apply classical protocols of error correction and privacy amplification and reducing to a negligible amount the eavesdropper's knowledge on the shared key. This second part of the cryptosystem is usually called classical communication phase.
Examples of QKD protocols based on the steps described above are BB84, B92 and BM92 \cite{BB84,B92,BBM92} that implement DV-QKD, and CV-QKD protocols like those introduced in References \cite{coh2002, 2way2006, 2way2016C}.

Previous QKD protocols suffer from the relative vulnerabilities connected to imperfections and trust-ability of devices used in  practical implementations. To overcome this difficulty, and potential security threats, an even more powerful approach to QKD has been introduced based on entanglement verification, and taking the name of Device-Independent (DI) QKD. In this approach the verification of violation of Bell inequalities is used to verify the presence of entanglement between the quantum signals shared between the parties. If entanglement is present then the parties will be in the position to share an unconditionally secure sequence of bits, ruling out any possibility for Eve to acquire information on the secret-key.  The  seminal work using entanglement to implement QKD has been proposed by Ekert in his 1991 work \cite{Ekert91}. After that many other works followed with refined security proofs \cite{pirandolaAQC}.

DI-QKD is the ultimate approach to establish unconditionally secure secret-key using quantum mechanics without having to specify the physical implementation of equipments or fixing many potential quantum hacking loopholes \cite{zhang2022device}.
However, DI-QKD is difficult to implement and its performance, on a practical scenario, are still limited \cite{pirandolaAQC} because it requires loophole-free Bell inequality violations, which necessitate high-quality entanglement among distant parties and near-perfect quantum detection, something current technologies cannot still provide in full \cite{zapatero2023advances}, or at least not under commonly accepted practicality assumptions.

In recent years also Measurement Device-Independent (MDI) QKD protocols have been proposed to implement overcome difficulties connected to the trustability of measurement devices \cite{braunsteinMDI,LoMDI,pir2016NatPhot}, and Twin-field QKD \cite{TFQKD} to overcome the point-to-point quantum secret-key capacity, set by the PLOB bound \cite{PLOBSKC}, and recover the single-repeater scaling of end-to-end quantum capacity \cite{RepPlob}, without the need to implement a full-scale quantum repeater. 

\textbf{4.2 Post-Quantum Cryptography:} The security of classical cryptographic primitives (e.g., RSA, Diffie–Hellman, etc.) depends on the hard problems of discrete arithmetic, prime factorization of integers, and elliptic-curve discrete logarithms. Sadly, these present-day cryptographic primitives based on such hard problems might theoretically be solvable in a brief span of time using the possible applications of quantum computers. The potential attacks performed by quantum algorithms posed on conventional cryptographic protocols, have promoted a sense of urgency in designing alternative schemes to mitigate quantum attacks. Such alternatives are generally characterized as post-quantum cryptography (PQC). These schemes can effectively deal with prevalent challenges triggered by quantum adversaries. The threat represented by the potential implementation nof fast quantum algorithm able to break the conventional algorithm used in our everyday life lead to an intense research activity on identifying candidates algorithm for the implementation and update of communication infrastructure able to resits to attacks performed to know quantum algorithms ~\cite{bernstein2017post}. The protocols developed in post-quantum cryptography were generally grouped into five types: code-based, hash-based, lattice-based, multivariate, and supersingular curve-elliptic isogeny schemes~\cite{kumar2022securing}.

NIST post-quantum cryptography standardization process~\cite{NISTPQC} is underway to identify the specific algorithm families and protocols to be considered secure under the potential threat of a quantum computer.

It is worth to notice that the ultimate counter-measures to preserve security and privacy of communication against \emph{quantum} eavesdroppers is quantum-key distribution (QKD), also against the possibility of \textit{``harvest now, decrypt later''} approach. In which case attackers store encrypted material until when advances on decryption technology (hardware or software) allows to decrypt the stored content. It is clear that in case of extremely sensitive data this may represent a threat to security that cannot be neglected, i.e., where data needs to remain confidentially protected for very long period of time.

\section{High-Scalability Quantum Computers}
Although quantum technology as a whole began in the 1980s, most scientists didn't see industrial quantum computers as feasible until the end of the 1990s \cite{gill2021quantum}. Several competitors, including academics and industrial engineers from throughout the world, have worked individually to construct the components of a robust quantum computer. Various potential material systems are being researched to design and implement quantum bits and gates. Analog and digital methods are the two most common ways to physically build a quantum computer. The preservation of qubit states owing to decoherence is a major obstacle to the building of error-free large quantum computers. The complexity of quantum circuits needed to tackle real-world issues could be substantial, leading to deleterious cumulative error rates, regardless of error rates attained below $1\%$ \cite{reed2012realization}. For this reason, the correction of quantum errors is currently a hot topic of academic interest. On October 23, 2019, Google Quantum AI and NASA announced a demonstration of quantum computation that would take a long time on any typical traditional computer \cite{arute2019quantum}. The successful resolution of a realistic everyday issue on a quantum computer is anticipated to necessitate much more research, despite the fact that this study accomplished an important step for the current batch of quantum computers. Importantly, IBM scientists demonstrated that identical computation can be executed far more efficiently on a conventional supercomputer \cite{pednault2019leveraging}.
 
\textbf{5.1 Super-fast Quantum Machines:} The ``quantum supremacy'' of quantum machines over conventional computers proves that the former can do very computationally intensive jobs on a conventional computer far more quickly. In the quantum world, ``quantum advantage'' is an additional important phrase. A more realistic concept would be ``quantum advantage'', which deals with solving a practical, real-world issue that cannot be effectively addressed on a traditional computer, as opposed to the theoretical ``quantum supremacy'' that would imply resolving a challenging issue on any conventional processor \cite{preskill2023quantum}. Quantum superiority has been shown, but finding real-world problems that quantum computers can effectively tackle remains unsolved mainly due to the decoherence of quantum bits. Most of the current generation of quantum computers is cumbersome and underpowered due to the materials used, which must be maintained at superconducting (extremely low) temperatures; yet, the promise of prospective commercial quantum computers is undeniable \cite{de2021materials}. The current popularity of traditional computers and their meteoric rise in the 1950s provide the impetus for the possible advantages of industrial quantum computers. Older classical computers were cumbersome and required constant cooling, just like modern quantum computers. We may theoretically expect strong commercial quantum systems to attain ``quantum advantage'' in the not-too-distant future, much as the  Artificial Intelligence (AI) concept began to take shape during the initial stages of traditional computing devices, even though these machines couldn't have possibly handled the computations needed for AI \cite{daley2022practical}.

\textbf{5.2 Quantum Computers for Business World: } The goal of cryptanalysis is to uncover the hidden features of a database. To decipher encrypted messages, it is necessary to bypass their cryptographic safeguards \cite{kumar2022securing}. To encrypt data transmission with banking as well as additional network nodes, one common method is the RSA algorithm \cite{biswas2023analysis}. If a massively error-corrected quantum machine could be built, the quantum technique that Shor created in 1994 might theoretically crack the operational RSA encryption. This highlights the necessity for the development of post-quantum algorithms for encryption that are resilient against commercial quantum computers. These days, many major companies place a premium on effective search strategies and the ability to effectively filter through massive datasets. When compared to conventional algorithms in terms of query complexity, Grover's optimum quantum algorithm from 1996 may significantly accelerate search across huge amounts of data \cite{grover1996fast}. Modern database management systems like Oracle aren't robust enough to handle Grover's algorithm in the actual world; hence, new software that mimics Oracle's functionality in the quantum realm is required \cite{gill2022quantum}. Approximation, rather than precision, is used to solve equations in many branches of computer research, including numerical weather forecasting and mathematical chemistry. In a weather/climate forecasting model, for instance, the parameterisation approaches employed to simulate sub-grid-level phenomena are a direct result of the computing limitations \cite{singh2022quantum}. The propagation of inaccuracies in the system of equation solutions brought about by these approximate parameterisations can have an impact on the decision-making process. Using commercially available quantum machines, we may be able to solve the equations exactly. In order to enhance the existing production process, which has a significant carbon footprint, this might shed light on how various chemicals are used to manufacture fertilisers. Quantum mechanical phenomena, chemical engineering, transpiration, superconductors, and magnetics may all be exploited with the help of commercial quantum machines \cite{gill2022quantum}. Investigation at the concept level has begun utilising accessible, comparatively less powerful quantum computers, even though a scalable industrial quantum computer has yet to be developed and may require substantial additional research. A beryllium hydride molecule was recently simulated on a seven-qubit quantum processor by IBM \cite{kandala2017hardware}. In the future, a number of applications are anticipated to gain popularity, including real-time consumer and transportation modelling, medical diagnosis by rapid database comparison, and power supply and demand balancing. However, the creation of commercial quantum computers will inevitably expose several other sectors and applications to risks, including communications, vital infrastructure, banking, the distributed ledger (blockchain), and cryptocurrencies, among others.

\textbf{5.3 Commercial Quantum Computing Infrastructure Specifications:} More than a hundred laboratories, including those associated with the government and universities, are working together on a global scale to develop, build, and monitor qubit systems \cite{de2021materials}. Production of commercial quantum machines is now underway at several big firms and a plethora of aspiring start-ups. In addition to creating quantum bits and gates, a commercial quantum machine would also need complex classical management and wiring, including cooling systems, user interfaces, networks, data storage capacities, and electromagnetic fields.

\textbf{5.4 Scalable Commercial Quantum Computing Manufacturing Challenges:} The biggest technical problem that needs to be solved before an industrial-grade quantum machine can be fully functional is noise or decoherence, which makes quantum processing mistakes (destroys entanglement of qubits) and stops quantum computing benefits. Until a stable qubit can be realised, its starting state must be established, and gates and networks must also be developed. Even though photons maintain their coherent state for an extended period of time, it is difficult to construct quantum circuits using them. Companies like IBM, Google, Rigetti, and others are building quantum machines using quantum circuits based on superconductivity. Unfortunately, there is still a need to develop strategies for error correction or moderation due to the poor fidelity of these qubits, especially in two-qubit operations. If a quantum circuit utilises five or fewer qubits, we can build and operate it on IBM's five-qubit cloud processor, which was made publicly available in 2016. In addition to their newly revealed 433-qubit quantum computer, IBM now provides cloud usage of quantum machines with up to 65 qubits.

\textbf{5.5 Presently Accessible Infrastructure:} In 2016, IBM unveiled its five-qubit IBM Quantum Experience quantum computer \cite{sisodia2020comparison}. Along with the system's release, a user manual and an interactive chat were made available. Rights to engage via quantum assembly language, a user-friendly interface, and a simulation extension were among the many features introduced to the IBM Quantum Experience later in 2017 \cite{piattini2021toward}. After that, IBM released Qiskit, a tool that enhanced quantum processor coding. In addition, they established the quantum awards program and created a system with sixteen qubits. Superconducting qubits housed in a dilution refrigerator constitute the hardware of IBM's quantum computers. The quantum composer is the name of the application's user interface (GUI) that consumers engage with. When writing quantum assembly code, quantum composer is the tool of choice. Quantum experiments and algorithms may be more easily developed with the help of the Graphical User Interface (GUI). One can also choose to use a simulator instead of a real Quantum Processing Unit (QPU). To run quantum computations through their paces, Rigetti Computing provides a Forest framework as a cloud-based quantum computing utility. A quantum processor from Forest has over 36 qubits, and it is possible to utilize Python to do hybridized conventional and quantum computations. The European cloud computing provider QuTech offers the quantum platform Quantum Inspire as part of its service offering. Without investing in or constructing a physical quantum computer, users can access the processing power of quantum algorithms using cloud-based quantum computing platforms.

\section{Widening the Debate: New Trends and Potential Challenges}

\begin{figure*}
	\centering
	\includegraphics[width=.9\textwidth]{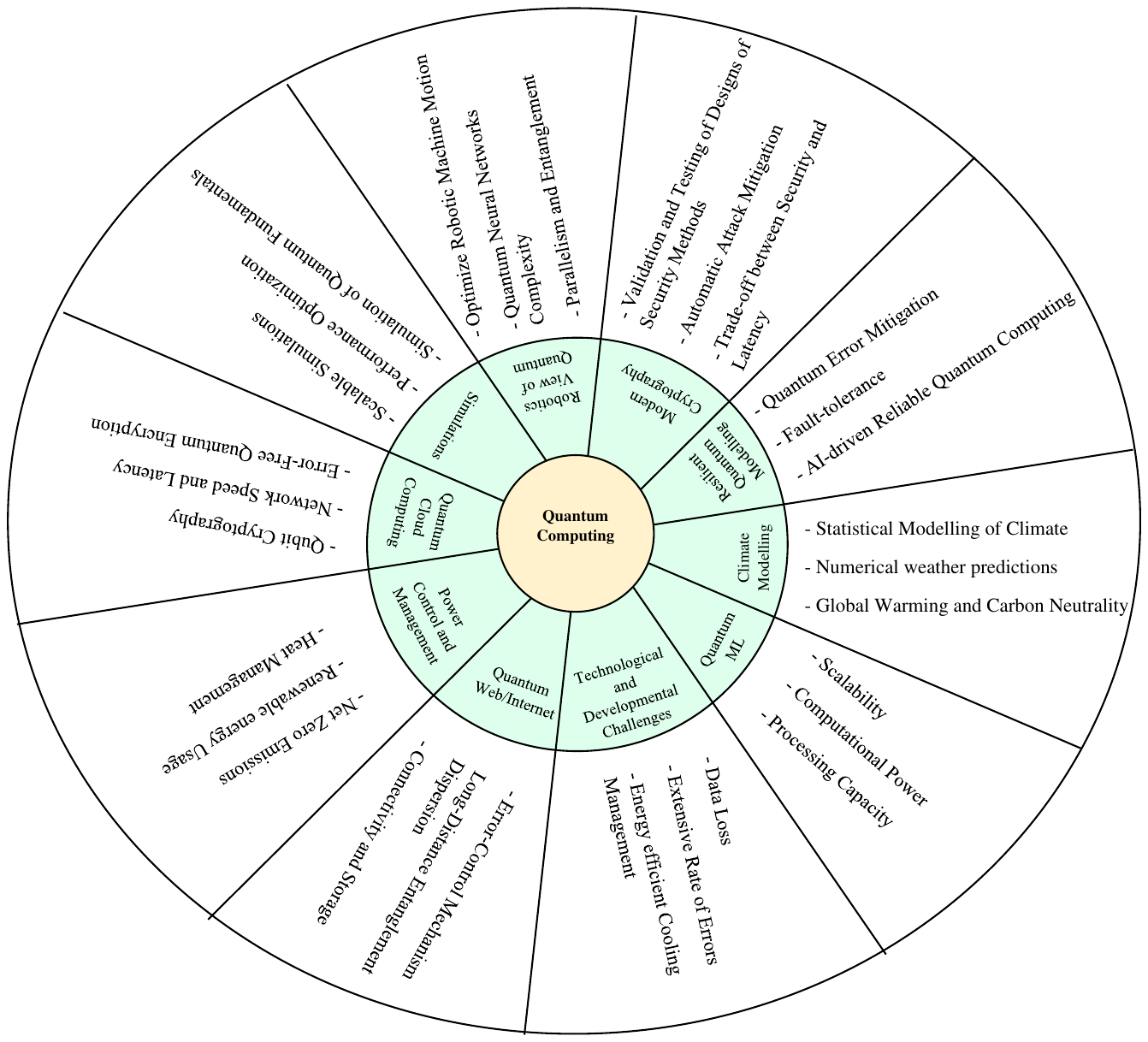}
	\caption [Caption for LOF] {New Trends and Potential Challenges in Quantum Computing.}
	\label{Trends}
\end{figure*}

In light of the current study, we have been able to pinpoint a number of topics in quantum computing that are still being investigated. Simulating complicated quantum processes has been the focus of much study, and post-quantum cryptography is now at its pinnacle. Fig.~\ref{Trends} summarises the main findings and recommendations that can be utilised by future researchers to further quantum computing research. In the realm of quantum technology, new fields of study are taking shape, including automation, handling energy, computer security, decentralised quantum computing, complicated mathematical chemistry and drug design \cite{preskill2023quantum}. It could take over a decade for these domains to fully implement quantum computing when they are first introduced. People have unrealistically high hopes for isothermal quantum computing, quantum management, and quantum security. Assuming they fall within the ambit of quantum computing, their development is anticipated to take short time \cite{kumar2022quantum}. There has been an excess of optimism around several areas of quantum technology, including the Internet, error-corrected quantum technology, digital information exploration, quantum-aided AI, and quantum-based satellite communications \cite{subramanian2022artificial}. We have uncovered several unanswered questions and potential avenues for further study, all of which are subject to ongoing investigation on a worldwide scale.

\textbf{6.1 Technological and Developmental Challenges: }The primary problem with quantum technology is its vulnerability, which arises from two main sources: 1) the fact that  qubits have a very short coherence period (which is very qubit technology dependent) since, due to their superconductivity, they lose their data extremely often. 2) Developing a quantum computer with minimal errors is challenging since quantum processes are unreliable because of the relatively substantial rate of errors needing a huge number of qubits for error handling. Additionally, error correction in quantum technology is far more challenging than in conventional computing due to the following reasons:  (a) quantum errors are ongoing (including the two magnitudes and stages), (b) it is not possible to replicate unknown quantum states, and (c) evaluation may degrade a quantum state and erase the information in qubits.  A large number of physical qubits are needed to execute a quantum algorithm successfully; this necessitates a tight and constant link between the classical structure and the quantum device, which in turn creates a massive control burden. Additionally, the connection and overhead costs increase the complexity of the run-time control, design, and installation for quantum computing processes. At the moment, the qubit count serves as a measure of quantum computing equipment's computational capacity. However, this metric is off by a significant margin, and it raises questions about the viability of supercomputer-level quantum machines with over a thousand qubits. Qubit design necessitates an efficient cooling component to manage heat, which AI-driven systems may be able to do. This increases scalability and allows for the solution of dynamically scaled, tricky issues.

\textbf{6.2 Resilient and Sustainable Quantum Modelling: } Since the actual application of quantum error mitigation remains a matter of wide debate, it is difficult to achieve trustworthy and fault-tolerant quantum computers. The sensitive nature of quantum states necessitates operating bits at extremely cold temperatures and requires high precision manufacturing \cite{pirandola2016physics}. Accurately measuring the full quantum state is similarly difficult, making verification a difficult task. When compared to conventional computing, the likelihood of calculation mistakes is much higher. Quantum structures cannot function properly without a reliable method of error correction. In order to facilitate better verification of exact manufacturing restrictions, further reevaluation of quantum communication infrastructure is required. However, due to strict tolerances and the need to prevent using poorly positioned qubits to minimise error, testing qubits after manufacture is a challenging task. To achieve sufficient reliability to enable sustained quantum computation, iterative error mitigation is required \cite{de2022software}. To provide trustworthy service in the years to come, state-of-the-art AI/ML-based methods may be utilised for automatic error identification and rectification on the fly \cite{gill2022ai}. Nonetheless, it results in additional expenses for training AI/ML  methods \cite{walia2023ai}. 

However, improving the dependability of computations does not only pass through more reliable hardware. In their seminal work, Avižienis et al. \cite{Avizienis200411} defines a taxonomy of dependable computing reporting applicable countermeasures at hardware and software levels. Software techniques to improve traditional computations and to tolerate hardware faults are nowadays a common practice in computer engineering. The challenges are to extend such software engineering practice to pursue highly dependable quantum programs \cite{7167244}; on the other hand, correct-by-construction is still a valid aim of software engineering, also applied to quantum computing: the application to quantum of model-driven engineering, formal modelling, advanced verification and validation techniques are other future challenges to deal with \cite{piattini2021toward}.

\textbf{6.3 Quantum ML \& QAI:} The use of principal component analysis, quantifying vectors, classifiers, regression, and stochastic modelling are common tools used by machine learning scientists. Using quantum computers to manage massive datasets with gadgets ranging from 100 to 1000 qubits may increase the effectiveness and scalability of AI methods. Additionally, by rapidly creating and evaluating certain statistical distributions, including training in conventional and quantum generative algorithms, quantum computers might pique the curiosity of the field of machine learning. As a result of the increasing amount of inputs (the number of participants) for quantum recommendation algorithms, it is becoming increasingly challenging to complete the task in a timely manner. Millions of qubits are required to deal with big datasets and present demand. By supplying computational power and other machine learning tasks, hybrid quantum-classical algorithms can overcome this challenge \cite{gill2022quantum}. Limited qubit connection and increased decoherence in the qubits caused by the device's intrinsic noise are two additional important problems. The use of sophisticated AI/ML can improve scalability and provide additional processing capacity to manage massive amounts of data produced by different Internet of Things (IoT) gadgets \cite{singh2023edge}.

\textbf{6.4 Power Control and Management:} Modern supercomputers and cloud servers need a great deal of electrical power to tackle various issues, making managing energy a major difficulty. When performing a specific activity, quantum computers are anticipated to use less energy compared in comparison. However, a quantum computer could consistently do massive computations with less power, cutting costs and reducing greenhouse gases even more. It can find the best answer with the least amount of energy because its qubits can represent both zeros and ones simultaneously for superposition (though entanglement or interference is also needed for computation), in contrast to classical computers' usage of binary bits (0 or 1). Quantum processors use less power since they operate at a shallow temperature, and because they are superconducting and have no resistance, they don't generate any heat \cite{gill2022quantum}. The two halves of an integrated application are the extremely energetic and low-energy components. Classical computing uses the cloud to execute the low-energy part, whereas quantum computing handles the high-energy portion \cite{gill2022quantum}. Therefore, hybrid computing, which combines quantum and conventional computing, can address these types of challenges since it significantly reduces energy consumption and expenses. To address the most difficult business issues of the present, further research is required prior to using hybrid computing. Utilising AI, quantum computers are capable of improving processing speed, dependability, and confidentiality \cite{gill2022ai}. However, this comes at a cost—a tremendous quantity of energy is required to power them and manage their temperature with cooling devices. Renewable energy sources, in conjunction with brown power, will be able to provide the energy needs for such quantum computers in the decades to come. 

\textbf{6.5 Quantum Web/Internet:} The advent of the quantum Internet has greatly improved computing power and opened the door for novel forms of communication, paving the way for decentralised quantum computing. The usage of quantum mechanics principles introduces a number of difficulties in the development of the quantum Internet, the most significant of which are the prohibitions on replication, quantum measurement, teleportation, and entanglement. A basic premise of conventional computing—the error-control mechanism—is now completely irrelevant in the context of quantum computing. In order to build the quantum Internet, a radical change from the current classical approach to networking design is required  \cite{pirandola2016physics}. Furthermore, decoherence results from qubit interactions with their environments due to the fragility of qubits and the gradual loss of qubit-to-environment information  \cite{wehner2018quantum}. Quantum computing has additional difficulties with efficient data transformation due to long-distance entanglement dispersion. It will be more difficult in the eventual quantum Internet to save the specifics of processes executed, which is a major drawback of current quantum computing systems that rely on massive amounts of storage for processing and connectivity.

\textbf{6.6 The Robotics View of Quantum: } Robots employ Graphics Processing Units (GPUs) to tackle computationally heavy problems in industries like pharmaceuticals, logistics, encryption, and banking, whereby the addition of quantum computing may significantly accelerate computations. Robots powered by quantum technology may also use cloud-based quantum computing resources to address a variety of problems \cite{gill2022quantum}. Modern industrial robots with improved sensing capabilities, made possible by quantum computing, may detect many jet engine problems simultaneously \cite{de2021materials}. In addition, by making use of two essential aspects of quantum computing—parallelism and entanglement—quantum image processing aids in the optimal understanding of visual knowledge as well as the efficient preservation and management of image data. Robots powered by AI are solving a wide range of issues by mining graphs for hidden insights, but the complexity grows exponentially as data sets get larger. By utilising quantum random walks rather than graph search, quantum computing is able to decrease performance. In addition, quantum neural networks may improve machine activities and detect instances of joint friction and motion, two additional major kinematics concerns. This means they can handle mechanical and robotic movements as well. In addition, there is another difficult challenge that may be tackled using quantum algorithms: determining why there is a discrepancy between the predicted and observed behaviours. The potential applications of quantum-reinforced learning might optimise robotic machine motion by addressing issues like joint friction and instances of inertia.

\textbf{6.7 Simulations for Advanced Quantum Research:} In the near future, small-scale "quantum simulators" with 50–100 qubits of computing power may be accessible, allowing quantum computers to model complicated biological, physical, and chemical issues \cite{gill2022quantum}. To comprehend and utilise quantum technology, it is necessary to combine the knowledge of several experts with the essentials of conventional computing \cite{daley2022practical}. In addition, quantum simulators can mimic the natural system and solve complicated issues in a controlled environment, allowing researchers to study the interplay of several parameters—questions that would be impossible to accomplish using conventional or supercomputer systems. When developing quantum computers, simulators can make use of entanglement and superposition, two of their key features \cite{piattini2021toward}. To conduct large-sized and complicated operations connected to biology and chemistry with optimum outcomes, the scalability of simulations needs to be increased in the future.

\textbf{6.8 Modern Cryptography:} Cryptography is essential for the safety of Internet communication, embedded medical equipment, and services. However, once big quantum computers are available, they will compromise the several commonly employed cryptosystems. Cryptographic algorithms, often known as public-key algorithms, are referred to as post-quantum cryptography. With post-quantum cryptography, it is presumed that the assailant used a massive quantum computer to launch the assault, and these systems adapt to remain safe in this scenario \cite{kumar2022securing}. Authenticity and secrecy must be preserved in post-quantum cryptography in order to thwart various assaults. Generally speaking, six methods—symmetric key quantum resistance, code-based, hash-based, multifaceted, and lattice-based encryption—are the primary focus of post-quantum cryptography investigation. Finding the correct places to include agility is a different issue within post-quantum cryptography. So, it's important to design ulterior systems with the ability to anticipate potential security issues. In addition, new automated techniques for fault detection and adaptive fixation during runtime are required for the validation and testing of designs \cite{mikkelsen2007optically}. A further unresolved issue is the necessity to integrate agility into old programs in order to reconfigure existing equipment with security protocols. Research in the future should focus on developing code-based systems that are more secure and produce results with less latency. As a result, research into the relative merits of latency, security, and data throughput is essential. Our goal is to achieve high processing and communication speeds while maintaining security. There has to be the formalisation of several standards in order to accommodate the shift to post-quantum cryptography in applications that operate in real-time. Understanding post-quantum method options is necessary for coordination with vital infrastructure, rescue services, mobile Internet financial services, and distance learning. Additionally, various methods can be chosen to hasten the transfer. 

\textbf{6.9 Statistical Modelling of Future Climate: } Improvements in computerised weather forecasting abilities occurred in the 1950s concurrently with the introduction of classical computers. Forecasts for the climate have come a long way in the years since, though. Though advancements in software and hardware have accelerated this trend, the use of bits, or 0s and 1s, as the building blocks of conventional computers has stymied progress. Highly powerful computers are constructed by stacking conventional computers to handle the massive amounts of computation that are needed. Every day, these supercomputers crunch numbers to predict what the planet's atmosphere, seas, and land will do. For practical uses in society, such as flood projections, metropolitan modelling, underground flow modelling, and related complicated tasks, today's advanced forecasts require significant improvements \cite{singh2022quantum}. The current state of computing power has impeded these advancements. The future global computer systems might be able to operate at significantly greater temporal and spatial detail if commercial quantum computers become feasible. Numerical weather forecasts using quantum computers require careful investigation. Since conventional computers' constraints generate inaccurate, high-resolution forecasts, numerical weather forecasting can benefit from quantum computing. With the processing capability of traditional computers being a constraint, the scientific objective is to solve complicated partial differential equations on the three-dimensional in natural spherical air and sea.

\textbf{6.10 Quantum Cloud Computing:} With the eventual widespread availability of robust quantum computers, unconditionally secured quantum cloud computing has the potential to play a significant role in a range of practical applications \cite{yang2023survey}. It could become considerably easier for the customer's work if there were a few strong quantum-computer nodes in the cloud. In order to transmit their work and related qubits, clients would have to interact with quantum servers using a quantum connection. There have been attempts to prove blind quantum computing through experimentation, in which quantum servers are unaware of the inputs, delegations, calculations, or outputs \cite{corcoles2019challenges}. The ubiquitous and potent quantum clusters have stymied these advancements. Methods for error-free quantum encryption, digital encryption basic concepts, and key distribution in a quantum cloud computing setting, as well as quantum approaches for gaining control in the cloud, are all covered in the following works: cryptographic verification of quantum computing, fault-tolerant secure quantum computations. Finally, in order to implement widespread quantum computing on a massive scale, research into a safe and effective quantum cloud computing platform is essential. Additionally, the quantum computing industry will benefit from using clouds as a means of storing, processing, and disseminating information \cite{piattini2021toward}. To overcome issues with network speed and latency that arise during the running of tiny activities in these systems, fog/edge computing is a viable solution \cite{walia2023ai}. The concept of blockchain may also be applied to the provision of reliable and safe services \cite{gill2021quantum}. 

\section{Summary of Findings, Take-Aways and Conclusions} 
There are several unanswered questions and some good ideas for where to go from here. To date, it has been unclear how to combine these performance features into a single quantum computing approach. In order to construct a quantum computer capable of concurrent activities, a quantum computing approach that enables quantum I/O to have all the required classified properties is important. A post-quantum cryptography system is developed to safeguard conventional cryptographic basics and protocols by using the computational power of a quantum computer, which can solve mathematical issues in milliseconds. In order to make symmetrical cryptography basics and algorithms more resistant to the widely-known quantum assaults, post-quantum cryptography was developed. Additionally, the difficulties in scaling up the number of qubits that have been actually realised thus far mean that modern commercial quantum computers have yet to be capable of replacing conventional supercomputers. It is uncertain when that may occur. There is currently no clear indication of when quantum computers will begin to supplant conventional computers in difficult tasks, despite the fact that the next decade will be absolutely thrilling for industrial quantum computing. Even if quantum computing does become feasible, digital supercomputers will continue to exist as a complement to potential quantum computers. The question of how to effectively operate an algorithm with quantum properties is a critical one for designers. There is significant control overhead due to the high number of physical qubits that are necessary, which in turn require constant and tight communication between the classical substrate and the quantum device. Due to the ongoing issue of the correction of quantum errors, it is difficult to accomplish trustworthy and resilient quantum calculations. The sensitive nature of quantum states necessitates operating bits at extremely cold temperatures and precise manufacture. Additionally, using quantum computing to manage a massive dataset with an extensive number of gadgets (100–1000 qubits) might enhance the effectiveness and scalability of AI methods. To realistically apply hybrid computing (quantum and conventional computing) and tackle today's most difficult business challenges, further effort is required. The advent of quantum computing will have far-reaching benefits for many other areas, including computer security, biology, economics, and the production of new substances.

Finally, this article offers a vision and identifies various potential challenges on the topic of quantum computing. It has been found that entanglement and superposition, two quantum mechanics instances, are anticipated to be crucial for resolving computer issues. We discussed a number of quantum software methods and technologies, industrial quantum computers, and cryptography after quantum computers. Finally, we highlight a number of concerns that have yet to be solved, as well as promising new prospects for research and development in the field of quantum technology.

\section*{Acknowledgements}

Ji Liu acknowledges support from the DOE-SC Office of Advanced Scientific Computing Research AIDE-QC project under contract number DE-AC02-06CH11357. 

\bibliographystyle{ieeetr}

\bibliography{cas-refs}

\end{document}